# Bandwidth comparison of photonic crystal fibers and conventional single-mode fibers

M.D. Nielsen[1,2*], J.R. Folkenberg[1], N.A. Mortensen[1], and A. Bjarklev[2]

[1] *Crystal Fibre A/S, Blokken 84, DK-3460 Birkerød, Denmark*

[2] *COM, Technical University of Denmark,
DK-2800 Kongens Lyngby, Denmark*

*[*mdn@crystal-fibre.com](*mdn@crystal-fibre.com)

**Abstract:** We experimentally compare the optical bandwidth of a conventional single-mode fiber (SMF) with 3 different photonic crystal fibers (PCF) all optimized for visible applications. The spectral attenuation, single-turn bend loss, and mode-field diameters (MFD) are measured and the PCF is found to have a significantly larger bandwidth than the SMF for an identical MFD. It is shown how this advantage can be utilized for realizing a larger MFD for the PCF while maintaining a bending resistant fiber.

©2004 Optical Society of America

## References and Links

1. T.A. Birks, J.C. Knight, and P.St.J. Russel, "Endlessly single-mode photonic crystal fiber," Opt. Lett. **22**, 961-963 (1997).
2. N.A. Mortensen and J.R. Folkenberg, "Low-loss criterion and effective area considerations for photonic crystal fibers," J. Opt. A: Pure Appl. Opt. **5**, 163-167 (2003).
3. M.D. Nielsen, N.A. Mortensen, and J.R. Folkenberg, "Reduced microdeformation attenuation in large-mode-area photonic crystal fibers for visible applications," Opt. Lett. **28**, 1645-1647 (2003).
4. W.A. Gambling, D.N. Payne, and H. Matsumyra, "Cut-off frequency in radially inhomogeneous single-mode fibre," Electron. Lett. **13**, 139-140 (1977).
5. D. Marcuse, "Gaussian approximation of the fundamental modes of graded-index fibers," J. Opt. Soc. Am. **68**, 103-109 (1978).
6. N.A. Mortensen, J.R. Folkenberg, M.D. Nielsen, and K.P. Hansen, "Modal Cut-off and the V-parameter in Photonic Crystal Fibers," Opt. Lett. **28**, 1879-1881 (2003).
7. M.D. Nielsen, N.A. Mortensen, J.R. Folkenberg, and A. Bjarklev, "Mode-field radius of photonic crystal fibers expressed by the V-parameter," Opt. Lett. **28**, 2309-2311 (2003).
8. M.D. Nielsen and N.A. Mortensen, "Photonic crystal fiber design based on the V-parameter," Opt. Express **11**, 2762-2768 (2003), http://www.opticsexpress.org/abstract.cfm?URI=OPEX-11-21-2762
9. T. Sørensen, J. Broeng, A. Bjarklev, E. Knudsen, and S.E.B. Libori, "Macro-bending Loss Properties of Photonic Crystal Fibre," Electron. Lett., **37**, 287-289 (2001).
10. O. Humbach, H. Fabian, U. Grzesik, U. Haken, and W. Heitmann, "Analysis of OH absorption bands in synthetic silica," J. Non-Cryst. Solids **203**, 19-26 (1996).

## Introduction

In conventional single-mode fibers (SMF), the single-mode optical bandwidth is typically limited by a higher-order mode cutoff at short wavelengths and macro-bend loss at long wavelengths. The characteristics of the photonic crystal fiber (PCF) are fundamentally different from this picture. Most important is the fact that the PCF can be designed to be endlessly single-mode (ESM), a term first coined by Birks et al. [1] referring to the fact that no higher-order modes are supported regardless of the wavelength. The ESM property has the specious consequence that the waveguide can be scaled to an arbitrary dimension while remaining single mode. However, as the scale of the structure is increased, the susceptibility towards attenuation induced by variations in structural parameters as well as external perturbations such as bending increases [2,3] limiting the practical dimensions that can be realized.

Often the question of which fiber type is most bend insensitive is raised: A PCF or a conventional SMF? The problem when trying to answer this question is that it is not very precise because the fibers to be compared need to be identical in terms of other optical properties in order for the comparison to be meaningful. Most relevant is to compare fibers with the same mode-field diameter (MFD) since bend loss in general increases with increasing MFD. Even though one should have a PCF and a SMF with identical MFD at a given wavelength, another problem arises from the fact that the MFD of the two fibers vary quite differently as function of wavelength making comparison difficult. Also, the MFD of one or both of the fibers might not necessarily result from an optimal choice of parameters and, finally, the spectral dependency of the bend loss is quite different for the two types of fibers. In this paper, we address these issues and attempt to make the comparison of the optical bandwidth taking the mentioned difficulties above into account. For the comparison we use fibers that are single mode at visible wavelengths and focus on their applicability red, green, and blue light (RGB) applications. The considered PCFs are all made of pure silica with a triangular arrangement of air holes of diameter, $d$, pitch, $\Lambda$, and a core formed by omitting the central air hole of the structure.

**Theory**

A good way to illustrate the differences in the spectral properties of the SMF and the PCF is through the V-parameter. The V-parameter for the SMF, $V_{SMF}$, has traditionally been applied to derive the higher-order mode cutoff [4] as well as the MFD [5]. Recently, we suggested a V-parameter for the PCF [6], $V_{PCF}$, and showed that this also held the property of uniquely determining both the higher-order mode cutoff [6] as well as the MFD [7]. The expressions for $V_{SMF}$, and $V_{PCF}$, are given by:

$$V_{SMF} = 2\pi \frac{a}{\lambda} \sqrt{n_{co}^2 - n_{cl}^2} \tag{1a}$$

$$V_{PCF} = 2\pi \frac{\Lambda}{\lambda} \sqrt{n_{FM}^2(\lambda) - n_{FSM}^2(\lambda)} \tag{1b}$$

In Eq. (1a), $a$ is the core radius and $n_{co}$ and $n_{cl}$ are the refractive indices of the core and the cladding, respectively. In Eq. (1b), $n_{FM}(\lambda)$ and $n_{FSM}(\lambda)$ are the wavelength dependent effective indices of the fundamental mode (FM) and the fundamental space filling mode (FSM), respectively (see ref. [6] for a detailed discussion on $V_{PCF}$). In the expression for $V_{SMF}$, the refractive indices are taken to be constants and $V_{SMF}$, therefore, depends on the wavelength as $1/\lambda$. The spectral dependency of $V_{PCF}$ is very different from that of $V_{SMF}$, since the effective indices are strongly wavelength dependent resulting in the fact that the index difference counteracts the effect of the $1/\lambda$ dependency and results in $V_{PCF} \to V_0$ for $\lambda \to 0$, where $V_0$ is a constant dependent on $d/\Lambda$ [8]. It is, thus, the decreasing effective index difference with decreasing wavelength that limits the number of modes and also has the effect that bend loss is observed at short wavelengths for the PCF [9]. From a MFD point of view, the increasing index difference as function of the wavelength ensures a close to constant strength of the guiding resulting in a MFD that can be almost constant over a broad wavelength range [7]. This is in contrast to conventional fibers, where the constant index difference becomes insufficient, when the wavelength is increased, causing the mode to expand until guiding is lost and, thereby, limiting the bandwidth at longer wavelengths.

When attempting to compare two different types of fibers as in the case of the SMF and the PCF, ensuring identical MFD are not sufficient. As an example, a PCF with a MFD of 10 µm at an operating wavelength of 1µm can both be realized with the parameters $d/\Lambda = 0.19$, $\Lambda = 5$ µm and with $d/\Lambda = 0.45$, $\Lambda = 8$µm. In the first case, $V_{PCF} = 1.0$, whereas the other example yields $V_{PCF} = 3.1$, and since a high $V_{PCF}$ value is preferred from a robustness point of view [8], these two designs will have a very different bending loss properties. Basing a comparison on PCFs with relative small values of $d/\Lambda$, therefore, holds limited relevance. Thus, in order to ensure a fair comparison, both fibers should be designed and operated where they are most robust, i.e., close to cutoff, and it should, furthermore, be insured that the MFDs are identical.

**Experimental**

The experimental investigation is based on a commercially available conventional SMF and 3 different PCFs. The SMF has a Ge-doped core region, a numerical aperture of 0.13, a MFD of 3.5 µm +/-0.5 µm, a cladding diameter of 125 µm, and a long term minimum bend radius of 13 mm. The PCFs all have a cladding diameter of 125 µm, and the structural parameters are listed in table 1.

Table 1. Characteristics of the tested PCFs

| ID | $\Lambda$ [µm] | $d/\Lambda$ |
|---|---|---|
| LMA-5 | 2.9 | 0.44 |
| LMA-8 | 5.6 | 0.49 |
| LMA-11 | 7.0 | 0.44 |

The spectral attenuation characteristics of the investigated fibers were measured using a white light source and the cutback technique. In Fig. 1, the attenuation spectra from 400 nm to 1700 nm for the conventional SMF and the LMA-5 PCF are shown. The attenuation spectra of LMA-8 and LMA-11 are very similar to that of the LMA-5 and, therefore, left out for the sake of simplicity.

For the conventional SMF (red curve) a sharp peak is observed at 430 nm originating from the higher-order mode cutoff. At a wavelength of around 820 nm the fiber no longer guides and a steep loss edge is observed. For the LMA-5 PCF (black curve), no cutoff is observed (the peaks at 1380 nm, 1245 nm, and 945 nm results from OH contamination [10]), and the fiber, thus, guides a single mode in the entire spectral range investigated. Although the attenuation level of the LMA-5 PCF is slightly higher than what can be realized, it is actually lower than or comparable to that of the conventional SMF for wavelengths lower than around 600 nm. This is due to the benefit of the pure-silica core, which does not suffer from attenuation bands from defect centers at short wavelengths to the same degree as Ge-doped silica. When comparing the curves shown in Fig. 1, a significantly larger single-mode bandwidth is apparently available for the PCF than for the conventional SMF. In order to check, if this large bandwidth comes at a price, the bend loss properties and the MFDs are compared.

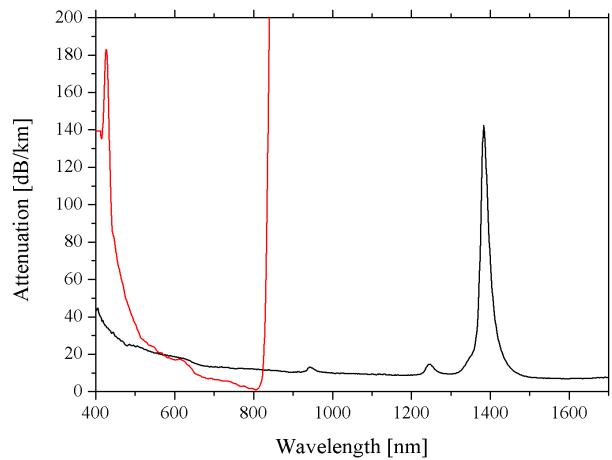

Figure 1 – The red and the black curve show the measured attenuation spectra of the conventional SMF and the LMA-5 PCF, respectively. The peak at 430 nm on the red curve indicates the higher order mode cutoff.

The bend loss properties of the investigated fibers were characterized by measuring the induced attenuation from 360° bends (single-turn bends) while varying bend radius, $R$. First, white light was coupled into the fiber and a reference spectrum was recorded with an optical spectrum analyzer, while ensuring $R > 80$ mm. The fiber was then given a single turn around a cylinder and a new transmission spectrum was recorded. This procedure was repeated for cylinders with $R$ from 10 to 80 mm (10 mm increments) and the attenuation was taken relative to the initial reference spectrum. The shortest bend radius of 10 mm is below the specified long term minimum bend radius for the conventional SMF and, therefore, not a value, which should be considered for most practical situations. It is, however, useful in order to test how far from the operational limit a fiber with good properties at $R = 20$ mm is. For this investigation the spectral range from 400 nm to 1000 nm was chosen since the conventional SMF, based on Fig. 1, will not operate at longer wavelengths.

In Fig. 2, the measured attenuation spectra are shown in panels A to D. Panel A shows the measurements for the conventional SMF. The shifting peaks in the spectral region between 400 and 450 nm results from bending induced shifting of the cutoff wavelength of the higher-order mode. It is, furthermore, observed, how the single-mode operating bandwidth decreases from the long wavelength side. At the smallest tested bend radius of 10 mm, the bandwidth extends from approximately 430 nm to 700 nm.

Panel B shows the corresponding measurements for the LMA-5 PCF and in this case no influence of the bending is observed. The LMA-5 PCF is, therefore, more robust than it needs to be for RGB applications and this unutilized potential could, therefore, preferably be used for realizing a larger MFD in the case where improved power handling is an issue. When increasing the structural scale by switching to the LMA-8 PCF, the sensitivity towards bending in increased as shown in Panel C. For the smallest applied bending radius of 10 mm, light is completely lost at wavelengths just below 500 nm. However, increasing the bend radius to 20 mm again results in the entire bandwidth being available with only a small indication of attenuation levels around 1 dB at 400 nm.

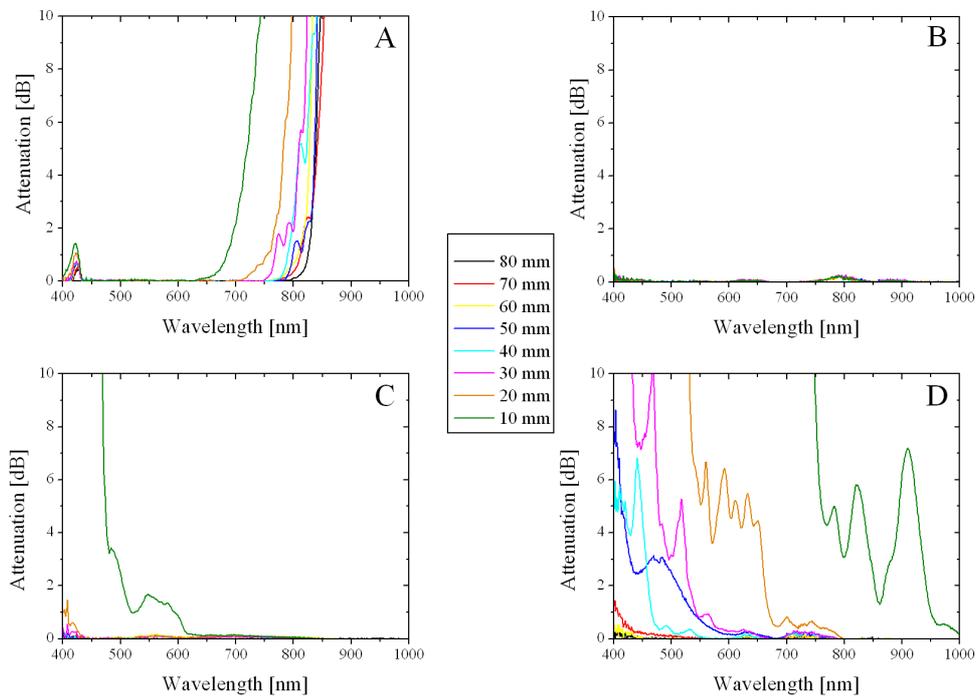

Figure 2 – Attenuation spectra from 400 nm – 1000 nm for single-turn bends with diameters as specified in the legend. Panels A, B, C, and D represent the conventional SMF, LMA-5, LMA-8, and LMA-11, respectively.

The measurements shown in panel C indicate that the structural parameters of the LMA-8 are in fact very close to being optimum in terms of having a fiber with the largest possible structure that is still robust towards any practical bend radius at any RGB wavelength. Bends of 10 mm are as mentioned not suited for long term operation. However, if the requirements for the robustness are less strict, further scaling of the structure is of course feasible. This point is illustrated in panel D, showing measurements for the LMA-11 PCF. In this case, the attenuation at 400 nm reaches 1 dB for R = 70 mm, and the fiber becomes dark for the shortest RBG wavelengths at R = 30 mm.

For investigation of the MFDs, we used a CCD camera to record an image of the fundamental mode at a number of wavelengths for each fiber. Light from light-emitting diodes (LEDs) at 470 nm, 525 nm, 570 nm, and 660 nm is coupled in and out of the fiber using microscope objectives. For each of the obtained images, a Gaussian function is fitted to the mode profile to yield a measure of the MFD. In order to calibrate the absolute scale of the profile, light is coupled into a short piece of PCF (∼10 cm), which is kept as straight as possible. By defocusing the coupling at the input, it is possible to guide light in the cladding region over this relatively short distance and to record the spatial distribution with the CCD camera. Since light is guided by the high-index silica regions of the cladding, the image will be an image of the fiber structure with a pattern of dark regions resembling the air holes. By measuring the period along such a line of dark regions, the pitch given in pixels can be determined. From an optical- or an electron micrograph, the pitch can afterwards accurately be determined and the absolute MFD of the near fields determined.

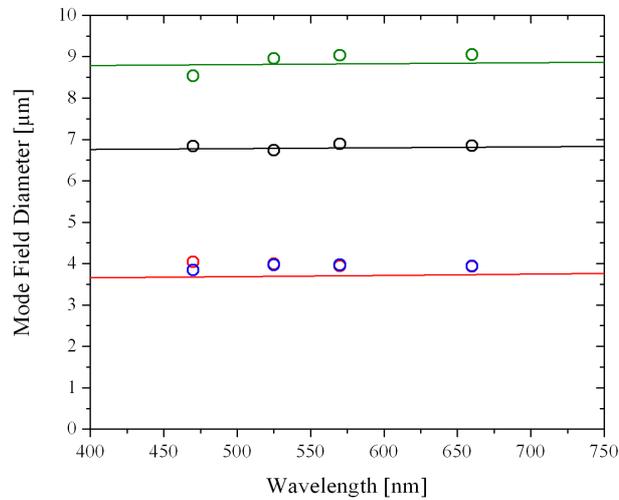

Figure 3: MFD data for the investigated fibers. The green, black, and red solid curves show the theoretical MFD calculated from the V-parameter of the LMA-11, LMA-8, and LMA-5, respectively. The green, black, red, and blue circles indicate measured values of the MFD for the LMA-11, LMA-8, LMA-5, and the conventional SMF, respectively.

In Fig. 3, measured and calculated MFD data are shown. The green, black, and red solid lines represent calculated data for LMA-11, LMA-8, and LMA-5, respectively. The calculations are based on the relations given in ref. [8] and the structural parameters listed in Table 1. The open green, black, red, and blue circles indicate measured data for the LMA-11, LMA-8, LMA-5, and the conventional SMF, respectively. Good agreement between calculated and measured data is observed showing the strength of the relatively simple expressions given in [8]. It is actually possible to accurately extract the MFD as function of wavelength from an image of the fiber cross section, which is in contrast to the conventional fiber. The MFDs of the LMA-11 and LMA-8 are close to 9.0 and 6.9 µm, respectively, while the MFDs of the LMA-5 and the conventional SMF are both very close to 3.9 µm at the inspected wavelengths. The measurements of the MFDs and the fact that both fibers are operated close to their cutoff wavelength, where robustness is optimum, make it possible to conclude that a direct comparison of the bandwidth from Fig. 1 and panels A and B from Fig. 2 can be justified.

**Conclusion:**

We have compared the optical bandwidth of a conventional SMF and 3 different PCFs intended for RGB applications. The MFDs of the LMA-5 PCF and the conventional SMF were identical at the inspected wavelengths. The PCF showed to be more robust towards bending at any of the investigated wavelengths from 400 nm to 1000 nm compared to the conventional SMF. The enhanced properties of the PCF are explained through the strong wavelength dependency of the effective index difference between the guided mode and the cladding modes and can be utilized for realizing a larger MFD with the benefit of improved power handling properties.

**Acknowledgements**

M.D. Nielsen acknowledges financial support from the Danish Academy of Technical Sciences.